\documentclass[aps,prb,showpacs,twocolumn,floats,epsfig]{revtex4}
\usepackage{amssymb}
\usepackage{amsbsy}
\usepackage{amsmath}             
\usepackage{epsfig}

\begin{document}

\title{One-dimensional itinerant ferromagnets with Heisenberg
symmetry and the ferromagnetic quantum critical point}

\author{K. Sengupta and Yong Baek Kim}

\affiliation{Department of Physics, University of Toronto,
             Toronto, Ontario M5S 1A7, Canada}

\date{\today}


\begin{abstract}

We study one-dimensional itinerant ferromagnets with Heisenberg
symmetry near a ferromagnetic quantum critical point. It is shown that
the Berry phase term arises in the effective action of itinerant
ferromagnets when the full SU(2) symmetry is present. We explicitly
demonstrate that dynamical critical exponent of the theory with the
Berry term is $z=2 +{\rm O}(\epsilon^2)$ in the sense of $\epsilon$
expansion, as previously discovered in the Ising limit.
It appears, however, that the universality class at the
interacting fixed point is not the same. We point out that even
though the critical theory in the Ising limit can be obtained by the
standard Hertz-Millis approach, the Heisenberg limit is expected to be
different. We also calculate the exact electron Green functions
$G(x,t=0)$ and $G(x=0,t)$ near the transition in a range of
temperature, which can be used for experimental signatures of the
associated critical points.

\end{abstract}


\maketitle

\section{Introduction}

Quantum phase transitions from a paramagnetic to a ferromagnetic
(FM) state in two and three dimensional itinerant electron system
have been studied extensively in the past \cite{hertz,andy1}. The
standard approach in studying such systems, first advocated by
Hertz \cite{hertz}, involves two steps. First, one decouples the
electron-electron interaction term in the Hamiltonian using
a Hubbard-Stratonovitch transformation. Next, the fermions are
integrated out to obtain an effective Landau-Ginzburg (LG) theory
in terms of the FM order parameter which provides an effective
description of the transition. However, it has been recently
pointed out that the procedure of integrating out the gapless
fermions in the above mentioned scheme might be tricky
\cite{kirk}. In particular, it might lead to singular terms in the
effective LG functional which can change the critical property of
the transition.

In contrast, very little attention has been paid to such transitions
in one dimension (1D), partly due to the Lieb-Mattis theorem
\cite{lm} which states that the ground states of one-dimensional
Hubbard models with nearest-neighbor hopping and density dependent
short-range interaction are always spin singlets. However, numerical
work carried out in Ref.\ \onlinecite{daul}, has shown that the
presence of a next-nearest-neighbor hopping term in the above
mentioned class of models can result in a FM ground state. The
properties of such 'FM Luttinger liquids' has been studied in detail
in Ref.\ \onlinecite{lorenz}. Recently, a bosonization analysis has
also been carried out to study the paramagnetic-FM  transition in 1D
Hubbard models with Ising symmetry \cite{yang1}. It is shown that in
$d=1$, the critical theory is below its upper critical dimension and
is governed by an interacting fixed point. A similar conclusion was
reached in Ref.\ \onlinecite{subir2} from analysis of rotor models
of ferromagnets.

In this work, we carry out a similar analysis which applies for
Heisenberg ferromagnets by explicitly keeping track of the full SU(2)
symmetry of the model. In particular, we show that at the Gaussian
level, the presence of a Berry phase term in the effective action of
such ferromagnets does not change the dynamical critical exponent $z$
of the transition: $z=2+{\rm O}(\epsilon^2)$ in the $\epsilon$
expansion sense.  We point out that in spite of the same value of the
dynamical critical exponent $z$ at the Gaussian level, the
universality class of this transition can be different from its Ising
counterpart. Our analysis also shows that contrary to the claim of
Ref.\ \onlinecite{yang1}, the critical theory for Ising symmetry can
be obtained by the standard Hertz-Millis procedure and in 1D such a
procedure is identical to the bosonization approach for Ising FM
quantum critical points. However, for the SU(2) case, it appears that
the critical theory may not be described by the standard Hertz-Millis
procedure.  We also consider the effect of the FM critical fluctuations
on the electron Green function at finite temperature. We
obtain an exact expression for the electron Green functions
$G(x=0,t)$ and $G(x,t=0)$ in a temperature range, which can be
accessed experimentally by measuring the tunneling density of states
(DOS) and the momentum distribution function respectively.

The organization of the paper is as follows. In the next section, we
obtain the effective action for 1D quantum ferromagnets with
Heisenberg symmetry at the Gaussian level and determine $z$ by
analyzing this action. Then we discuss the interacting fixed
point in the $\epsilon$-expansion sense and the issue of universality
of the Ising and Heisenberg cases in the light of possible quartic couplings.
This is followed by Sec.\ \ref{egsec}, where we
obtain the electron Green's function at finite temperature.
We summarize and discuss our results in Sec.\ \ref{dis}.

\section{Critical theory in the Heisenberg limit}
\label{berrysec}

We first derive the effective action for the
critical theory by incorporating the transverse spin
fluctuations in Sec.\ \ref{derivs}. This is followed by the analysis
of the effective action to derive the dynamical critical exponent $z$
in Sec.\ \ref{analysis}.

\subsection{Derivation of the effective action}
\label{derivs}

We start from a one dimensional $t$-$J$ model given by
\begin{eqnarray}
{\mathcal H} &=& - \sum_{<ij>} \left[ \sum_{\sigma} t
\left(\psi_{i\sigma}^{\dagger} \psi_{j\sigma} +{\rm h.c.}\right) + J
{\bf S}_i \cdot {\bf S}_j \right] \label{heisen1}
\end{eqnarray}
where $\psi_{i\sigma}$ represents an electron at site $i$ with spin
$\sigma$, and ${\bf S}$ denotes the electron spin operators
$S_{i a} = \psi_i^{\dagger} \sigma_{a} \psi_i$ where $\sigma_a$ are the
usual Pauli matrices. The corresponding action can be written as
\begin{eqnarray}
S &=& \int dx \int_0^{\beta} d\tau \left[\psi^{\dagger}(x,\tau)
\left(\partial_{\tau}-\mu\right)\psi(x,\tau) +{\mathcal H} \right]
\label{ac0}
\end{eqnarray}
where $\tau$ is the imaginary time and $\beta=1/k_B T$ denotes
inverse temperature. Note that $\psi^{\dagger} =
(\psi^{\dagger}_{\uparrow}, \psi^{\dagger}_{\downarrow})$ and $\psi
= (\psi_{\uparrow}, \psi_{\downarrow})^T$ are spinor representations
of the electron operators. We have switched to a continuous
representation of space and this amounts to replacing the last term
in Eq.\ \ref{heisen1} by $-\int dx dx' J(x-x') {\bf S}(x) \cdot {\bf
S}(x')$ where $J(q)=J \cos(q_x a)$, where $a$ is the lattice spacing
and we shall consider $J>0$ throughout as appropriate for studying a
ferromagnetic instability.

Next, following Schultz \cite{schulz1}, we introduce new fermionic
fields $\psi'$ defined as $\psi (x,\tau) = U (x,\tau)
\psi'(x,\tau)$. Here $U$ represent SU(2) rotation matrices such
that $U^{\dagger}(x,\tau) {\bf \sigma} \cdot {\bf n}(x,\tau)
U(x,\tau) = \sigma_z$.  The unitary matrices $U$ therefore correspond to
the rotations of the local spin quantization axis of the electrons
to ${\hat z}$. The action $S$, in terms of these new fields, becomes
\begin{eqnarray}
S &=& S_0 + S_J \nonumber\\
S_0 &=& \int dx d\tau \ \psi'^{\dagger}
\Big [ \partial_{\tau}-\mu- C_0
-2t \cos\left(-i\partial_x -C_x\right) \Big ] \psi'
\nonumber\\
S_J &=& - \int d\tau dx dx' \ J(x-x') S^z(x) S^z(x') \label{ac1}
\end{eqnarray}
where the fields $C_{\mu} = -i U^{\dagger} \partial_{\mu} U =
\sum_{a=x,y,z} \sigma_{a} \Omega_{\mu}^{a}/2 $ are
the SU(2) gauge fields which describes the spin fluctuations of
the system, $\mu=(\tau,x)$ refers to the time and space components
of the fields and $\partial_{\mu}= (i\partial_{\tau},\partial_x)$.
The gauge freedom here consists of rotation of the spin of the
$\psi'$ fields about the $\hat z$ axis: $\psi'\rightarrow
\psi'\exp\left(i\sigma_z \Phi \right)$. Such a rotation changes
$\Omega_{\mu}^z \rightarrow \Omega_{\mu}^z + \partial_{\mu} \Phi$
and therefore leaves the action invariant.

From now on we shall replace $\psi'$ by $\psi$ for notational
convenience. The fermion fields can be written in terms of the right
and the left movers: $\psi(x) = \psi_L(x) e^{-ik_Fx} + \psi_R(x)e^{i
k_F x}$, where $k_F$ is the Fermi wavevector. The right and the left
moving fermions have energy dispersion
\begin{eqnarray}
\epsilon_{\alpha}(k_x) &=& {\rm sgn}(\alpha)v_F k_x +\frac{k_x^2}{2m}
\end{eqnarray}
where $v_F=2ta\sin\left(k_F a\right)$ is the Fermi velocity and
$m=1/\left[4ta^2\cos\left(k_F a\right)\right]$ is the mass for the fermions
and ${\rm sgn}(\alpha)
 =+(-)$ for $\alpha=R(L)$. In most of our
analysis, we shall restrict ourselves to the linearized dispersion
relation for the fermions, which amounts to neglecting the
quadratic term in $\epsilon_{\alpha}(k_x)$. However, for deriving
an effective action for the spin-fluctuations we shall need to
retain the quadratic term, and hence it can not be neglected at
the outset \cite{nicolas1}.

In terms of these fields, the action reads
\begin{eqnarray}
S &=& S_0 +S_1 + S_2 + S_J \nonumber\\
S_0 &=& \int d^2k \sum_{\alpha} \psi_{\alpha}^{\dagger}(k)
G_0^{-1} \psi_{\alpha}(k)
\label{ac2a}\\
S_1 &=& \int d^2k \ d^2p  \sum_{\alpha}
\psi_{\alpha}^{\dagger}(k+p) \nonumber\\
&& \times \left[\frac{\partial G_0^{-1}}{\partial k_{\mu}} ,
C_{\mu}(p)\right]_{+} \psi_{\alpha}(k)
\label{ac2b} \\
S_2 &=& \int d^2k \ d^2p \sum_{\alpha}
\psi_{\alpha}^{\dagger}(k+p) \nonumber\\
&& \times \left[\frac{\partial^2 G_0^{-1}}{\partial k_{\mu}^2} ,
C_{\mu}^2(p)\right]_{+} \psi_{\alpha}(k)
\label{ac2c}\\
S_J &=& -\int d^2 q \sum_{\alpha \alpha'} J_{\alpha \alpha'}(q)
S_{\alpha}^z(q) S_{\alpha'}^z (-q) \label{ac2d}
\end{eqnarray}
Here $\int d^2 k$ is the shorthand notation for
$\beta^{-1}\sum_{\omega_n} \int dk_x/(2\pi)$, $k_\mu = (i\omega_n,
k_x)$, $[..]_+$ denotes anticommutator, $\alpha$ denotes the $R,L$
indices for the right(R) and left(L) moving fermions, $G^{-1}_0 =
i\omega_n -\epsilon_{\alpha}(k_x)$ is the free fermion propagator,
and $J_{\alpha \alpha'}(q)=J_{\alpha \alpha'}\cos(q_x)$.
We have neglected all $2k_F$ terms in $S$ since we shall be primarily
interested in studying the action near a ferromagnetic
instability. The matrix $J_{\alpha \alpha'}= J+\delta J
\delta_{\alpha \alpha'}$ has an additional $\delta J$ term in its
diagonal components, which is added to make the matrix $J$
invertible and will be set to zero at the end of the calculation.
The term $S_2$ (\ref{ac2c}) represents the diamagnetic term of the
SU(2) gauge fields $C_{\mu}$ and is known to be important for
deriving a low energy effective description of the spin
fluctuations. Note that to obtain this term, one needs to go
retain the quadratic term in $\epsilon_{\alpha}(k_x)$
\cite{nicolas1}.

We then carry out a Hubbard-Stratonovitch transformation to
decouple $S_J$:
\begin{eqnarray}
S &=& S_0 +S_1 + S_2 + S_{I}+ S_{\phi} \nonumber\\
S_{I} &=& \int d^2k \sum_{\alpha} i\phi_{\alpha}(k)  S^z_{\alpha}(-k)\\
S_{\phi} &=& - \frac{1}{4}\int d^2k \sum_{\alpha \alpha'} (J(q))^{-1}_{\alpha
\alpha'} \phi_{\alpha}(q) \phi_{\alpha'}(-q)
 \label{ac3}
\end{eqnarray}
Notice that the field $\phi$ does not correspond to the physical
spin-density fields, but their conjugate. However, for studying the
ferromagnetic instability, it is preferable to look at the physical
spin-density fields. We therefore introduce these fields $\rho$ via a
second Hubbard-Stratonovitch transformation which decouples
$S_{\phi}$:
\begin{eqnarray}
S &=& S_0 +S_1 + S_2 + S_{\rm int}+ S_{\rm HS} \label{fermlast} \nonumber\\
S_{\rm int} &=& \int d^2k \sum_{\alpha} i\phi_{\alpha}(k)
\left[S^z_{\alpha}(-k)
- \rho_{\alpha} (-k)\right]\label{ac44} \\
S_{\rm HS} &=& -\int d^2k \sum_{\alpha \alpha'} J_{\alpha
\alpha'}(q) \rho_{\alpha}(q) \rho_{\alpha'}(-q)  \label{ac4}
\end{eqnarray}
The next task is to integrate out the fermionic fields and obtain an
effective low-energy description of the system in terms of the fields
$C_{\mu}$, $\rho$ and $\phi$. This procedure turns out to be quite
subtle as discussed in Ref.\ \onlinecite{nicolas1}, but can be carried
out \cite{comment1}. The resulting effective action, to Gaussian order, reads
\begin{eqnarray}
S_{\rm eff} &=& S_{\rm NLSM} + S_{\rm Berry} + S[\rho,\phi] \\
S_{\rm NLSM} &=& \frac{N(0)}{4} \int d\tau dx
\left[ \left(\partial_{\tau}{\bf n}\right)^2
+ v_F^2 \left(\partial_x {\bf n}\right)^2 \right]
\label{nlsm}\\
S_{\rm Berry} &=& \int d^2 q \Omega_0^z(q) \sum_{\alpha} \rho_{\alpha}(-q) \\
S[\rho,\phi] &=& \int d^2q \sum_{\alpha \alpha'} \Bigg(
\frac{1}{2} \phi_{\alpha} \Pi_{\alpha \alpha'}^{-1}
\phi_{\alpha'}
-i \phi_{\alpha}(q) \rho_{\alpha'}(-q) \delta_{\alpha \alpha'} \nonumber\\
&& - J_{\alpha \alpha'}(q)
\rho_{\alpha}(q) \rho_{\alpha'}(-q)\Bigg)
\label{eff1}
\end{eqnarray}
where we have expressed $S_{\rm NLSM}$ in terms of the unit vector
field ${\bf n}$ using the identity \cite{nicolas1}: ${\rm
Tr}\left[C_{\mu}\right]^2 -{\rm Tr}\left[\sigma_z C_{\mu}\right]^2 =
\left(\partial_{\mu} {\bf n}\right)^2/4$. The Berry phase term $S_{\rm
Berry}$, as is well known, can not be expressed in terms of ${\bf n}$
vector since $\Omega_0^z$ depends on gauge choice of the SU(2) fields
\cite{subir1}. The polarization tensor $\Pi_{\alpha \alpha'}$ in
$S[\rho,\phi]$ is given by
\begin{eqnarray}
\Pi_{\alpha \alpha'}^{-1}(q) = \delta_{\alpha \alpha'} N(0)
\frac{{\rm sgn}(\alpha) v_F q} {-i q_{0n} +{\rm sgn}(\alpha) v_F q}
\label{pol}
\end{eqnarray}
Within the linearized dispersion and in one dimension Eq.\
\ref{eff1} is known to be exact \cite{Kopietz}.

We now introduce the spin-density fields in the symmetric(S) and
the antisymmetric(A) channels $\rho_{S(A)}= \left(\rho_R
+(-)\rho_L\right)$, integrate out the auxiliary fields
$\phi_{\alpha}$ from Eq.\ \ref{eff1}.  Since the ferromagnetic
instability corresponds to the instability of the spin-density
fields in the symmetric channel, we also integrate out the
$\rho_A$ to obtain an effective action for $\rho_S$
\begin{eqnarray}
S_{\rm eff} &=& S_{\rm NLSM} + S_{\rm Berry} + S[\rho_S]
\label{sfinal} \\
S_{\rm Berry} &=& \int d^2 q \Omega_0^z(q) \rho_{S}(-q)
\label{berry} \\
S[\rho_S] &=& \int d^2q \rho_S(q) {\mathcal S}_S (q) \rho_S(-q)
\nonumber\\
&=& - \int d^2q \rho_S(q)
\left( \frac{q_0^2}{v_F^2 q^2} - r -b q^2 \right)
\rho_S(-q) \nonumber\\
\label{critical}
\end{eqnarray}
where $r=(1-JN(0))$, $b=JN(0)a^2/2$, and $a$ is the lattice spacing.
Here we have Wick-rotated back to real frequency and expanded
$J(q)\simeq J(1-q_x^2 a^2/2)$ . Notice that the ferromagnetic
instability in Eq.\ \ref{critical} is signaled by $r=0$ or $JN(0)=1$
which agrees with the usual Stoner criteria for ferromagnets.

We would like to stress that, unlike the Ising case studied in Ref.\
\onlinecite{yang1}, our final effective action (Eq.\ \ref{sfinal})
is not only $S[\rho_S]$, but also contains $S_{\rm NLSM}$ and
$S_{\rm Berry}$. This is crucial for a correct description of the
transverse spin fluctuations which are natural consequence of the
SU(2) symmetry of the problem. This point is best illustrated by
studying the action (Eq.\ \ref{sfinal}) inside the FM phase. Here
the spin-density fields are gapped out and $\left<\rho_S\right>=M$,
where $M$ is the magnetization of the system. In the FM phase, the
low energy theory is therefore described by $S_{\rm NLSM}$ and
$S_{\rm Berry}$. Note that the presence of the Berry phase term
is crucial for the $k_x^2$ dispersion of the spin waves, as pointed out
in Ref.\ \onlinecite{wen} in the context of 2D ferromagnets. A similar analysis
in this line for the Ising limit would yield Eq.\ \ref{critical} as the
final effective action which is exactly the same result obtained in
Ref.\ \onlinecite{yang1}.

We also note that a similar procedure can be carried out in
higher dimension. In that case, due to the presence of a continuous
isotropic Fermi surface, instead of two Fermi points, the index
$\alpha$ becomes the momentum $k_{\parallel}$ along the Fermi
surface and the sum over $\alpha$ has to be replaced by an integral
over $k_{\parallel}$. This changes the $q_0^2/(v_F^2 q^2)$ term in
${\mathcal S}_S(q)$ (Eq.\ \ref{critical}) to $|q_0|/(v_f q)$ and
immediately leads to $z=3$. This result can also be alternatively
obtained by finite dimensional bosonization technique
\cite{Kopietz}.

\subsection{Analysis of the effective action and critical theory}
\label{analysis}

We now analyze the critical theory for 1D. Our starting point is
Eq.\ \ref{sfinal} derived in the previous subsection. From Eq.\
\ref{nlsm}, the fluctuations of the ${\bf n}$ field are gapless in
the Heisenberg limit, it is not a priori clear whether the coupling
of $\rho_S$ to these fluctuations changes the universality class of
the transition. To check this, we aim to integrate out the
transverse fluctuation modes to obtain the final effective action in
terms of the $\rho_S$ fields. To this end, we rewrite the action
$S_{\rm NLSM}$ and $S_{\rm Berry}$ using the CP(1) representation.
In this representation, the unit vector field ${\bf n}$ is
represented by two complex fields $(z_1,z_2)$ with the constraint
$|z_1|^2 +|z_2|^2=1$. Following Refs.\ \onlinecite{Witten,adda}, one
can write
\begin{eqnarray}
S_{\rm NLSM} &=& \int dt \ dx \sum_{j=1,2} |\left(\partial_{\mu}
- i A_{\mu} \right)z_j|^2
+\lambda (|z_j|^2 -1) \nonumber \\
 \label{cp1}
\end{eqnarray}
where the U(1)gauge fields $A_{\mu}$ are given by
\begin{eqnarray}
A_{\mu}&=& \Omega_{\mu}^z = i \sum_{j}
\left[z_j^{\ast}\partial_{\mu} z_j - (\partial_{\mu} z_j^{\ast})
z_j\right]
\label{gauge1}
\end{eqnarray}
The field $\lambda$ is a Lagrange multiplier field used to implement the
constraint of unit norm for the CP(1) fields. Note that in this
representation, $S_{\rm Berry}$ can be written as
\begin{eqnarray}
S_{\rm Berry} &=& \int d^2q  A_{0}(q) \rho_S(-q)
\label{cp2}
\end{eqnarray}
and has the convenient interpretation of the action describing
coupling of a U(1) scalar potential to matter field.

To describe the Berry term in a gauge invariant fashion, we now define a
spin-current $j_x^z$ which satisfies the continuity relation
\begin{eqnarray}
\partial_t \rho_s + \partial_x j_x^z &=& \partial_{\mu} J_{\mu}(x)=0
\label{cont}
\end{eqnarray}
and write the Berry term as
\begin{eqnarray}
 S_{\rm Berry} &=& \int d^2q  A_{\mu}(q) J_{\mu}(-q)
\label{cpb}
\end{eqnarray}
Although this procedure seems ad-hoc, the additional term in $S_{\rm
Berry}$ introduced here could easily be obtained rigorously from our
starting action (Eq.\ \ref{ac1}) if we carried out the analysis in
the presence of a spin-current interaction term $ H_{\rm current} =
-\int dx \int dx'\, K(x-x'){\bf j}(x) \cdot {\bf j}(x')$ where
$j_x^{\alpha} = \psi^{\dagger} \sigma^{\alpha} \partial_x \psi(x)$
is the spin-current operator. Such a current-current interaction
term, although always present in principle, is usually neglected
since it is very small (O($v_F^2/c^2$)) compared to $S_J$. The only
relevant information, apart from the continuity condition
(\ref{cont}), that we shall need for our analysis here is that the
spin-current correlator in the symmetric channel is given by
\cite{comment2}
\begin{eqnarray}
{\mathcal K}_S(q) &=& - v_F^2 \left( \frac{q_0^2}{v_F^2 q^2}-1
+ K \cos(qa) N(0) \right) \nonumber\\
&& \simeq - v_F^2 \left( \frac{q_0^2}{v_F^2 q^2} - r' \right)
\label{sccor} \nonumber\\
\end{eqnarray}
where $r'= 1-KN(0)$. We therefore simply use this information and
avoid carrying out a detailed analysis involving $H_{\rm current}$.
Note that the propagator ${\mathcal K}_S$ is always massive near the
critical point. Hence it is not necessary to retain the $q^2$ term in
its expression.

At this stage, we find that to understand the effect of $S_{\rm
Berry}$ on the ferromagnetic quantum critical point, we need to
obtain an effective action for the gauge fields $A_{\mu}$. This
involves integrating out the fields $z$ and $\lambda$ from $S_{\rm
NLSM}$. It is well known that such an analysis can be reliably
carried out for CP($N-1$) theories only in the large $N$ limit
\cite{Witten,adda}. However, it is conjectured in Ref.\
\onlinecite{Witten} that the qualitative results hold even for
$N=2$. We therefore follow the large $N$ analysis of Refs.\
\onlinecite{Witten,adda} to obtain
\begin{eqnarray}
S_{\rm eff} &=& S_{\rm gauge} + S_{\rm Berry} +S_{\rm matter} \\
S_{\rm gauge} &=& -\int d^2 p A_{\mu}(p) \left(p^2\delta_{\mu \nu}
- p_{\mu} p_{\nu} \right) A_{\nu}(-p) \nonumber\\
&=& \int d^2p A_{\mu}(p) G_{\mu \nu}^{\rm bare}(p) A_{\nu}(-p)
\label{gaug1}
\end{eqnarray}
where $p_{\mu}=(p_0,v_F p_x)$. Notice that the propagator $G^{\rm
bare}$ is not invertible.  This problem can be easily taken care
of by simply adding a gauge fixing term, as is customary for the
photon propagator in QED. However, we shall not need to invert the
bare propagator at any stage of our analysis, and hence retain the
form of $G^{\rm bare}$ as given in Eq.\ \ref{gaug1}. The action
$S_{\rm matter}$ has now to be supplemented by a quadratic term in
the spin current $j_x^z$ which involves
\begin{eqnarray}
S_{\rm matter} &=& \int d^2 q \Big[ \rho_S(q) {\mathcal S}_S(q) \rho_S(-q)
\nonumber\\
&& + j_x^z(q) {\mathcal K}_S(q) j_x^z(-q) \Big]
\label{matterac}
\end{eqnarray}

The final step is to integrate out the gauge fields. To do this, we
first integrate out the matter field $J_{\mu}$ and obtain an effective
dressed propagator $G^{\rm dressed}$ for the gauge fields $A_{\mu}$
\begin{eqnarray}
G^{\rm dressed}(k) &=& -\left(\begin{array} {cc} v_F^2 k_x^2
+\frac{1}{4{\mathcal S}_S(k)} & \omega v_F k_x  \cr
\omega v_F k_x & \omega^2
+ \frac{1}{4{\mathcal K}_S(k)}
\end{array} \right) \nonumber\\
\end{eqnarray}
We then replace $G^{\rm bare}$ by $G^{\rm dressed}$ in Eq.\
\ref{gaug1}, and integrate out the gauge fields to obtain the
contribution $\delta S$ of the gauge fields to the effective
action for the matter fields $J_{\mu}$ \cite{comment3}
\begin{eqnarray}
S'_{\rm eff} &=& S_{\rm matter} + \delta S  \nonumber\\
\delta S &=& \int d^2 J_{\mu}(p) (\delta P)_{\mu \nu} (p)
J_{\nu}(-p) \label{corr1} \\
\delta P (k) &=& \frac{{\mathcal K}_S(k){\mathcal S}_S(k)}{{\mathcal
K}_S(k)\omega^2 +v_F^2 k_x^2
{\mathcal S}_S(k)^2 + 1/4}  \nonumber\\
&& \times \left(\begin{array} {cc} \omega^2 +
\frac{1}{4{\mathcal K}(k)} &
-\omega v_F k_x  \cr -\omega v_F k_x & v_F^2 k_x^2
+ \frac{1}{4{\mathcal S}_S(k)}
\end{array} \right)
\end{eqnarray}
Note that the term $\delta S$ couples the $\rho_S$ and $j_x^z$
fields. To obtain the effective action for the $\rho_S$ fields, we
integrate out the $j_x^z$ fields and evaluate the resulting propagator
at the critical point $r=0$ and retain the lowest order terms in
$\omega$ and $k$. After some straightforward algebra, one obtains the
correction term $\delta S'[\rho_S]$
\begin{eqnarray}
\delta S'[\rho_S] \simeq - \int d^2 k \rho_S(k)
\left(\frac{\omega^2}{v_F^2 k_x^2} - k_x^2 + ...\right) \rho_S(-k)
\label{correc1}
\end{eqnarray}
where the ellipsis represent terms which are higher order in
$\omega$ and $k_x$. So we conclude that the effect of the Berry term
is merely to renormalize the coefficients of the existing terms in
$S[\rho_S]$. The critical theory obtained by integrating out the
transverse fluctuation at the Gaussian level thus has $z=2$. By the
usual $\epsilon$ expansion argument, we expect this analysis to give
correct value of $z$ to ${\rm O}(\epsilon^2)$.

\begin{figure}
\centerline{\psfig{file=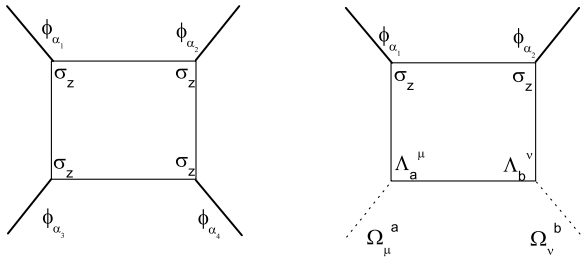,width=1.0\linewidth,angle=0}}
\caption{Representative diagrams for quartic coupling terms for
Heisenberg theory. The left diagram is also present for Ising
ferromagnets and generates the usual $\rho_S^4$ interaction term.
The diagram on the right side generates quartic couplings between
$\rho_S$ and ${\bf n}$ fields (see discussion in text), and has no
analog in Ising ferromagnets. The internal lines involve Fermion
Green functions and the vertices represent either $\sigma_z$ or
$\Lambda^{\mu}_a =  \sigma_a \partial G_0^{-1}/\partial k_{\mu}$ as
shown. Both the diagrams are nonzero only if the Fermion dispersion
has a non-zero curvature.} \label{fig1}
\end{figure}

Next we comment on the universality class of the transition. To
determine the universality class, since we are below the upper
critical dimension of the theory, we need to retain the possible
fourth order diagrams while integrating out fermions in Eq.\
\ref{fermlast}. Two such representative diagrams are shown in Fig.\
\ref{fig1}. The diagram in the left panel is present for the Ising
ferromagnets studied in Ref.\ \onlinecite{yang1}, while that in the
right panel is a consequence of the Heisenberg symmetry of the
problem. After some algebra \cite{Kopietz}, one can show that the
diagram in the left panel of Fig.\ \ref{fig1} generates Ising type
contributions $ \sim \int dx dt \rho_S^4$ while the diagram in the
right panel generates, for example, quartic couplings
$ \sim \int dx dt \rho_S^2 (\partial_x {\bf n})^2$ (apart from other
similar quartic and higher order terms). The former type of terms
are exactly those that are expected at quartic order for the Ising
ferromagnets, either from a bosonization analysis \cite{yang1} or
Hertz-Millis approach \cite{hertz,andy1}. Therefore, our analysis
shows that the standard Hertz-Millis procedure in 1D, gives exactly
the same critical theory as obtained in Ref.\ \onlinecite{yang1} by
bosonization for the Ising case.

The latter type of diagrams are absent in the Ising model studied in
Ref.\ \onlinecite{yang1}. In the FM phases, where the amplitude mode
is gapped, these terms represents a trivial renormalization of the
parameters such as spin-wave stiffness of $S_{\rm NLSM}$ (Eq.\
\ref{nlsm}). Near the transition, they generate quartic coupling
between the amplitude and the spin-wave modes. Notice that these
terms are different from the standard ${\bf m}^4$ quartic terms
expected from the usual Hertz-Millis analysis of a Heisenberg magnet
with a vector order parameter. We expect such quartic coupling terms
to change the universality class of the transition from that of
Ising ferromagnets. A determination of the universality class would
therefore require a detailed analysis of Eq.\ \ref{sfinal}
supplemented with all such possible relevant quartic coupling terms.
In particular, this requires a consideration of both coupled
longitudinal and transverse modes at the same footing. In this work, we
shall be content with pointing out the necessity of such an analysis.

The quartic terms mentioned above makes the Gaussian fixed point,
described by Eq.\ \ref{sfinal}, unstable. However, all such
diagrams, including the Ising type quartic terms, are non-zero only
when the fermionic dispersion has a finite curvature
\cite{yang1,Kopietz}. Thus, above a temperature scale set by this
curvature, the properties of the system is expected to be well
described by the Gaussian theory described by Eq.\ \ref{sfinal}. We
discuss this point in more details in the next section.

\section{Electron Green's function }
\label{egsec}

It is well known that the presence of a quantum critical point shows
up in experimentally accessible quantities such as tunneling DOS and
the momentum distribution function of the electrons. To obtain these
quantities one needs to estimate the effect of critical fluctuations
on the electron Green function. However, the fixed point concerned
here is an interacting one. This means to compute the electron Green
function at $T=0$ near the critical point, one needs to solve the
problem of electrons coupled to $interacting$ bosons. For the Ising
ferromagnets these bosons can be described by an action
\begin{eqnarray}
S &=& \int dx d\tau \Big[\phi \left(-\partial_{\tau}^2 - r
 \partial_x^2 + b \partial_x^4 \right) \phi +
u (\partial_x \phi)^4 \Big], \label{rgac}
\end{eqnarray}
where $u$ is determined by the curvature of the fermionic
dispersion. Obtaining the electron Green function coupled with these
interacting bosons, even for the Ising ferromagnets, is a difficult
task. However, it might be still be possible to compute the
properties of Green function at finite temperature using the
Gaussian fixed point, if the RG flow, which can be computed for
Ising ferromagnets from Eq.\ \ref{rgac}, generated by temperature
takes us to a regime where $b q_T^2 > r(T) \gg u^{\ast}$, where
$q_T=\sqrt{T/T_F} \,k_F/ \left[1-r(T)\right]^{1/4}$ is the thermal
wavevector and $u^{\ast}$ is the value of the coupling $u$ at the
interacting fixed point.

To explain the last sentence a little better, we consider the RG
flow of the action $S$ (Eq.\ \ref{rgac}) at finite temperature, to
one loop. As noted in Ref.\ \onlinecite{yang1}, the flow of $u$ and
$b$ are negligible as $r$ flows away from the fixed point according
to
\begin{eqnarray}
\frac{dr}{dl}= 2r + \frac{3u}{\pi}
\left(\Lambda^2-\frac{r}{2}\right)\coth\left(\frac{\Lambda^2}{2T}\right)
\label{rgf}
\end{eqnarray}
where $T=T_0 \exp(zl)$ is the scaled temperature. Therefore we may
envisage that when $\epsilon$ is small, there will be a temperature
$T_1$, above which $r(T)\gg u^{\ast}$, so that the effect of
interaction can be neglected. The key question is then whether at
and above $T_1$, $r(T)$ is small compared to the $\partial_x^4$ term
in $S$ (Eq. \ \ref{rgac}). A simple estimate shows that for this to
happen, we need to be below a temperature $T_2$ given by
\begin{eqnarray}
\frac{T_2}{T_F} &=& \frac{r(T_2)}{\sqrt{1-r(T_2)}}
\label{t2estimate}
\end{eqnarray}
For $T>T_2$, $r$ can not be neglected and the behavior of the
system is similar to that of a Luttinger liquid at finite
temperature.

If indeed $T_1 < T_2$, there exists a finite window where the effect
of the critical fluctuations on the electron Green function can be
calculated neglecting the effect of interaction. We refrain from
estimating $T_1$ and $T_2$ since they need the knowledge of precise
value of $u^{\ast}$ and is therefore non-universal in the RG sense.
Instead, in the rest of this section, we shall assume that such a
window exists and compute the behavior of the electron Green
function for $T_1<T<T_2$. Note that such an analysis, which requires
only properties of the Gaussian fixed point, also holds for the
Heisenberg ferromagnets, since the Gaussian action is the same in
both cases. The question of existence of such a window of course has
to be separately investigated for the Heisenberg ferromagnets. Such
an investigation requires a detailed analysis of quartic couplings
for the Heisenberg theory and is beyond the scope of the present
work. However since all such quartic terms depends on the curvature
of the Fermionic dispersion, we expect such a window to exist also
for the Heisenberg ferromagnets also when the curvature is small.

To calculate the Green function for the electron we start from the
action
\begin{eqnarray}
S &=& S_0 + S_1 \nonumber\\
S_0 &=& \sum_{\alpha}
\int d^2q d^2k \ \psi^{\dagger}_{\alpha}(k+q) \nonumber\\
&& \times \left (G_0^{-1} + i \sigma_z \phi_{\alpha}(q) \right)
\psi_{\alpha}(k) \nonumber\\
S_1 &=& \sum_{\alpha} d^2q \phi_{\alpha}(q) F_{\rm RPA}^{-1} (q)
\phi_{\alpha}(-q)
\label{sa}
\end{eqnarray}
where
$\phi$ denotes the bosonic
fields with the propagator $F_{\rm RPA}$ given by
\begin{eqnarray}
F_{\rm RPA}(q) &=& \frac{J(q_x)}{1+J(q_x)\sum_{\alpha}
{\tilde \Pi}_{\alpha \alpha}^{-1}(q)}
\label{rpaint}
\end{eqnarray}
where ${\tilde \Pi}_{\alpha \alpha}^{-1} = N (T) \alpha v_F q_x /\left(i\Omega_n
-\alpha v_F q_x \right)$ is the finite temperature generalization of
$\Pi_{\alpha \alpha'}^{-1}(q)$ in Eq.\ \ref{pol} for $q_x/k_F \ll 1$,
and $N(T) = \sum_k \partial f(\epsilon)/\partial (\epsilon) \simeq
N(0)$ is the free fermion DOS.  Note that the starting action given by
Eq.\ \ref{sa} can be easily obtained from Eqs.\ \ref{ac44} and
\ref{ac4} with the propagator for the $\rho(q)$ fields replaced by
$F_{\rm RPA}$ \cite{comment4}.

To obtain the Green's function of the electrons, we proceed following the
analysis carried out in Ref.\ \onlinecite{Kopietz}. The equation for the Green
function is given by
\begin{eqnarray}
\delta(k-k') &=&
\sum_{k_1} \left[ G^{-1}_0 (k) \delta(k-k_1)\delta_{\sigma \sigma'}
\right. \nonumber\\
&& \left. + i (\sigma_z)_{\sigma \sigma'} \phi_{\alpha}(k-k_1) \right]
G_{\alpha}^{\sigma \sigma'} (k,k')
\label{green1}
\end{eqnarray}
which in 1D can be solved exactly using Schwinger ansatz
\cite{Schwinger,Kopietz}. After some straightforward manipulations,
we get, $G^{\sigma \sigma'} (x,\tau)
=\delta_{\sigma \sigma'} G(x,\tau)$, where
\begin{eqnarray}
G(x,\tau) &=& \sum_{\alpha}e^{i\alpha k_F x} G_{\alpha}(x,\tau)
\nonumber\\
G_{\alpha}(x,\tau) &=& G_{0\alpha}(x,\tau)
\exp\left[Q (\alpha x,\tau)\right]
\label{greenf} \\
G_{0 \alpha} (x,\tau) &=& \int d^2k \ e^{i \left(k_x x-\omega_n
\tau\right)}  \frac{1}{i\omega_n -\alpha v_F k_x} \nonumber\\
&=& \left(\frac{-i}{2\pi}\right) \frac {\xi^{-1}}{\sinh\left[
\left(\alpha x - iv_F \tau \right)/\xi \right]}
\label{fgreen}\\
Q (\alpha x,\tau) &=& \int d^2 q \
F_{\rm RPA}(q) \frac{ 1-\cos\left(q_x x-
\Omega_n \tau \right)}{\left(i\Omega_n -
\alpha v_F q_x\right)^2}. \label{qfunc}\nonumber\\
\end{eqnarray}
Here $\omega_n$ ($\Omega_n$) denote fermionic (bosonic) frequencies
satisfying the usual anti-periodic (periodic) boundary conditions,
$\xi =\hbar v_F/k_B T $ is the thermal correlation length, and one
needs to make the standard Wick rotation $\tau= it$ to obtain the
Green functions in real time.

To proceed further, we note that experimentally accessible
quantities such as the momentum distribution $n_{\alpha}(k)$ or the
tunneling DOS $\rho(\omega)$, do not probe the full Green function
$G_{\alpha}(x,\tau)$, but only $G_{\alpha} (0,\tau)$ and
$G_{\alpha}(x,0)$, since
\begin{eqnarray}
n_{\alpha}(k_x) &=& \int
\frac{dk_x}{2\pi} \exp (ik_x x ) G_{\alpha}(x,0), \label{mdist} \\
\rho (\omega) &=& \int dt {\rm Im}\left[\exp(-i\omega t )
G_{\alpha}(0,t)\right].
\label{tdos}
\end{eqnarray}
Therefore we resort to the simpler task of computing $G(x,0)$ and
$G(0,\tau)$. The frequency sums in the expressions of $Q(x,0)$ and
$Q(0,\tau)$ (Eq.\ \ref{qfunc}) can now be evaluated in a
straightforward manner. We express the final result in terms of the
crossover temperature $T_1$ and the corresponding length scale
$\xi_1 = \hbar v_F/k_B T_1 = k_F^{-1} T_F/T_1$:
\begin{eqnarray}
Q(x,0) &=& \exp \Bigg(-\int^{\infty}_{q'_c} dq'
\frac{1-\cos(q'x')}{q'} \nonumber\\
&& \times \Bigg[ [1+\gamma(q')] \coth\left(\frac{\mu(q')
q'}{2s}\right) \nonumber\\
&& -\coth\left(\frac{q'}{2s}\right) \Bigg] \Bigg)
\label{q1exp} \\
Q(0,t) &=&  \exp \Bigg(-\int^{\infty}_{q'_c} dq' \Bigg[
\Bigg \{\frac{1-\cos(q'\mu(q')t')}{q'} \,  \nonumber\\
&& \times \left[1+\gamma(q')\right] \coth\left(\frac{\mu(q') q'}{2
s}\right) \nonumber\\
&& -\frac{1-\cos(q't')}{q'} \coth\left(\frac{q'}{2 s}\right)
\Bigg\}\Bigg] \Bigg), \label{q2exp}
\end{eqnarray}
where all momenta, length, and time scales are expressed in terms
of $q'= q_x \xi_1$, $x'=x/\xi_1$, $s=T/T_1$ is the dimensionless
temperature, and $t'=v_F t/\xi_1$. Here $\mu(q') = {\tilde
v}_F(q')/v_F$ is the ratio of the renormalized and the bare Fermi
velocities and $\gamma(q')$ is a dimensionless parameter given by
\begin{eqnarray}
\gamma(q')&=& \frac{1+\mu(q')^2}{2\mu(q')}-1 \nonumber\\
\mu(q') &=& \sqrt{r + b'q'^2} \quad \quad b'=b/\xi_1^{2}
\label{coeffc}
\end{eqnarray}
and $q'_c=q_c \xi_1$ is the dimensionless infrared cutoff. Note
that in our notation, the non-interacting limit corresponds to
$\mu(q)=1=r$.

Before looking at the Green function in its full generality, it is
instructive to check that it reproduces the well- known result for
the Luttinger liquid Green function when $b'=0$ and $r\ne 0$. In
this limit the integrals in Eq.\ \ref{q1exp} and \ref{q2exp} can be
calculated analytically and one gets at $T=T_1$,
\begin{eqnarray}
G_{\alpha}^{\rm LL}(x',0) &=& \left(\frac{-i}{2\pi}\right)
\left(\frac{\mu(0)/\xi_1}{\sinh(\alpha x/\xi_1)}\right) \nonumber\\
&& \times \left(\frac{\mu(0)/q'_c \xi_1}{\sinh \left(
|x'|\mu(0)/\xi_1\right)}\right)^{\gamma(0)} \nonumber\\
G_{\alpha}^{\rm LL}(0,t') &=& G_{\alpha}(\alpha x'=t',0).
\label{llg}
\end{eqnarray}
From Eq.\ \ref{llg}, one can identify $\gamma(0)$ to be the
anomalous dimension \cite{Kopietz}. Note that, as is well known for
Luttinger liquids, the Green functions $G(x,0)$ and $G(0,t)$ has
identical behavior. This is a consequence of the linear dispersion
of the bosonic density fluctuations.
\begin{figure}
\centerline{\psfig{file=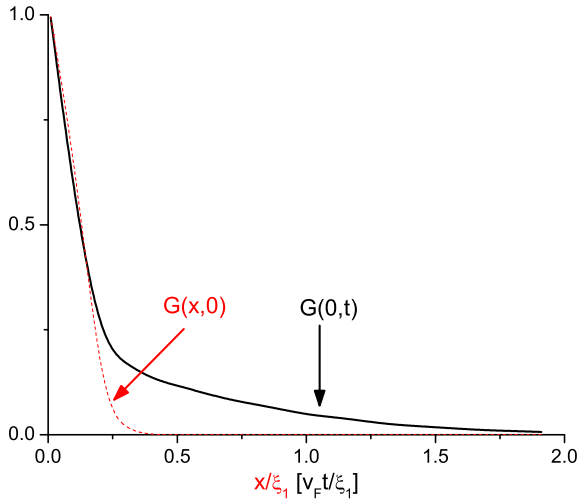,width=1.0\linewidth,angle=0}}
\caption{Plot of $G(x,0)$ (dashed line) and $G(0,t)$ (solid line)
near the critical point ($r=0$) at $T=T_1$. The decay of $G(0,t)$ is
slower than $G(x,0)$ as discussed in the text.} \label{fig2}
\end{figure}
As we approach the critical point, $r \rightarrow 0$, and
consequently one must retain the $b'$ term, in the expression of
$\gamma(q')$. When $b'\ne 0$, it is not possible to evaluate $Q$
analytically at finite temperature. However, since the electrons now
see the critical bosonic fluctuations with dispersions $\omega
\simeq q_x^2$, we expect the Green functions $G_{+}(x,0)$ and
$G(0,t)$ to have different behavior. A plot of the Green functions
shown in Figs.\ \ref{fig2}, confirms the above qualitative
discussion. The characteristic scale of decay for $G(x,0)$ is seen
to be much shorter compared to $G(0,t)$. This feature arises from
the presence of critical fluctuations with $\omega \sim q_x^2$
dispersion ( in contrast to $\omega \sim q_x $ fluctuations in
standard Luttinger liquid ) and is therefore a signature of the
quantum critical point at finite temperature. We expect that a
measurement of $\rho(\omega)$ and $n(k_x)$ will probe this behavior.

\section{Discussion}
\label{dis}

In this work, we have shown, by explicitly keeping track of the full
SU(2) symmetry of the problem, that in 1D the Luttinger liquid to FM
transition for Heisenberg ferromagnets has $z=2+{\rm O}(\epsilon^2)$
in an $\epsilon$ expansion sense. The analysis done here was carried
out at a Gaussian level, which is exact in 1D as long as the
dispersion of the Fermions are linear. The curvature in Fermionic
dispersion introduces quartic coupling terms beyond the Gaussian
approximation. Although we have not carried out a full analysis of
the problem with such coupling terms in the presence of the SU(2)
symmetry, we have identified the typical quartic terms and discuss
their consequences. In particular, we pointed out that for a
ferromagnet with Ising symmetry, our analysis reduces to the
standard Hertz-Millis theory. The Hertz-Millis approach is therefore
completely equivalent to the bosonization approach of Ref.\
\onlinecite{yang1} in 1D for Ising ferromagnets. In contrast, we
find that for the Heisenberg ferromagnets, our theory does not
reduce to that obtained by the traditional Hertz-Millis approach. We
identify that this is due to additional coupling terms between the
longitudinal and the transverse modes, shown in right panel of Fig.\
\ref{fig1}, which has no analog in the Ising case and are also
different from the usual expected ${\bf m}^4$ interaction terms that
one obtains using the Hertz-Millis approach. It will be interesting
to carry out our analysis in higher dimension and compare the
results with those of Ref.\ \onlinecite{kirk}, where a possible
problem with such ${\bf m}^4$ quartic terms in the usual
Hertz-Millis approach has been discussed from a different point of
view. A detailed analysis of this issue is left as a subject of
future work.

We have also obtained an exact expression for the electron Green
functions $G(x,0)$ and $G(0,t)$ for a range of finite
temperature. Although we have not obtained the expression of the Green
functions at the interacting fixed point, we hope that over a range of
temperature where the properties of the system is expected to be well
described by the Gaussian fixed point, $G(x,0)$ and $G(0,t)$ can be
probed by experimentally accessible quantities such as the momentum
distribution and tunneling DOS of the electrons.

This work was supported by NSERC of Canada, Canada Research Chair
Program, and Canadian Institute for Advanced Research. We thank Kedar
Damle, Dima Feldman, Subir Sachdev, and Kun Yang for their insightful
comments and Eugene Kim for early collaboration on a related project.


\begin{thebibliography}{99}
\bibitem{comment1} The procedure of obtaining a non-linear sigma model
for a gapped quasi one-dimensional system is outlined in Ref.\
\onlinecite{nicolas1}. We follow this procedure here and set the gap
$\Delta$ to zero at the end of the calculation. However, this also
means that the domain of validity of such the model, which is valid
for $E<\Delta$, shrinks to zero. This implies that the coefficients of
the model such as spin stiffness and susceptibility can not be
trusted. As far as our analysis here is concerned, it does not depend
on the precise values of these coefficients and is expected to hold as
long as the coefficients remain non-singular. We do not know of any
analysis which rigorously derives an effective model of spin
fluctuation for a metallic system.
\bibitem{comment2} Note that to obtain ${\mathcal K}_S$, one proceeds exactly the
same way as for ${\mathcal S}_S$.
We first obtain ${\mathcal K}_{\alpha \alpha'} = \left< j_{x \alpha}^z
j_{x \alpha'}^z \right>$ for $\alpha= R,L$, then switch over to the symmetric
and antisymmetric modes $j_{x\,S(A)}^z = j_{x\,R}^z +(-) j_{x\,L}$ and finally
integrate out the anti-symmetric modes to get ${\mathcal K}_S$.
\bibitem{comment3} The use of dressed, rather than the bare propagator of the
gauge field is crucial here and can be justified as an attempt
to take into account self-consistency of the problem. Of course, in general,
it is not necessary that this procedure will lead to self-consistency.
For the problem at hand, however, it can be checked a posteriori that the use
of dressed propagator for the gauge fields indeed gives us a self-consistent
solution in the infrared limit.
\bibitem{comment4} For Eq.\ \ref{rpaint} to hold, it is necessary that $J(q)$
is independent of the index $\alpha$. This point has been discussed in
details in Ref.\ \onlinecite{Kopietz}.
\end{thebibliography}
\end{document}